\documentclass{elsart4}

\usepackage{graphicx}
\usepackage{amssymb}
\usepackage[english,francais]{babel}

\newcommand {\apgt} {\ {\raise-.5ex\hbox{$\buildrel>\over\sim$}}\ }
\newcommand {\aplt} {\ {\raise-.5ex\hbox{$\buildrel<\over\sim$}}\ }

\newtheorem{e-proposition}[theorem]{Proposition}

\newtheorem{e-definition}[theorem]{Definition\rm}

\renewcommand{\theequation}{\arabic{equation}}
\def\og{\leavevmode\raise.3ex\hbox{$\scriptscriptstyle\langle\!\langle$~}}
\def\fg{\leavevmode\raise.3ex\hbox{~$\!\scriptscriptstyle\,\rangle\!\rangle$}}

\begin{document}
\centerline{Title of the dossier/Titre du dossier}
\begin{frontmatter}

\selectlanguage{english}
\title{The multiwavelength context in 2015 and beyond}

\author{Jochen Greiner},
\author{Arne Rau}
\address{Max-Planck-Institut f\"ur extraterrestrische Physik, 
  85748 Garching, Germany}

\begin{abstract}
  We collect  information as complete  as possible about  the upcoming
  telescopes and facilities relevant  to gamma-ray bursts, both in the
  electromagnetic bands as well as for non-electromagnetic messengers.
  We describe  the expected synergy  between these new  facilities and
  the  SVOM mission  and predict  possible  progress in  the field  of
  gamma-ray bursts over the next years.

{\it To cite this article: J. Greiner, A. Rau, C. R. Physique ?? (2011).}

\vskip 0.5\baselineskip

\selectlanguage{francais}
\noindent{\bf R\'esum\'e}
\vskip 0.5\baselineskip
\noindent
{\bf Le contexte multi-longueurs d\'onde en 2015 et au-del? }
Nous avons r\'euni des informations aussi compl\'etes que possible au   
 sujet des futurs t\'elescopes et dispositifs exp\'erimentaux pouvant   
 servir \`a l'\'etude des sursauts gamma, dans le domaine des ondes   
 electromagn\'etiques autant que par d'autres messagers. Nous anticipons   
 les progr\'es \`a attendre dans l'\'etude des sursauts gamma au cours des   
 années \`a venir, and nous pr\'esentons la synergie escompt\'ee entre ces   
 installations futures et la mission SVOM.

{\it Pour citer cet article~: J. Greiner, A. Rau, C. R. Physique ?? (2011).}

%Now keywords/mots-clM-is
 \vskip 0.5\baselineskip
\noindent{\small{\it Mots-cl\'es~:} Mot-cl\'e1~; Mot-cl\'e2~;
Mot-cl\'e3}}
\end{abstract}
\end{frontmatter}

%\selectlanguage{english}
% main text

\section{Introduction}

SVOM will  not only carry  an X-ray and  an optical telescope  for GRB
afterglow  studies,   but  the  ground-segment  is   also  planned  to
incorporate  an array of  wide field-of-view  optical cameras  for the
detection of  prompt optical emission, plus two  robotic 1-meter class
telescopes  for   immediate  optical/NIR  follow-up   observations  of
ECLAIR-discovered GRBs.  With these new features it  is interesting to
investigate which synergy SVOM will have with the other planned space-
and ground-based  facilities. This exercise is attempted  here, and we
want to stress two caveats right from the beginning: (1) the selection
of information has  been done as careful as  possible, but necessarily
remains subjective; (2) history  shows that most predictions of future
(GRB)  research results turned  out to  be wrong,  or were  dwarfed by
unpredicted  discoveries \cite{hur95,piro07}.  There  is no  reason to
believe that this will not happen again.

\section{Major successes in the field of GRB afterglows}
\label{}

The progress in  the Gamma-Ray Burst (GRB) field  over the last decade
and  prior  to  the launch  of  {\it  Fermi}  mostly occurred  in  our
understanding  of  the afterglow  emission  and  the GRB  surroundings
through  the   decisive  measurements  of  {\it   Swift}  and  enabled
ground-based follow-up  work. Classical observational  astronomy, from
radio to  X-ray energies, played a  vital role in this  progress as it
allowed  the   identification  of  GRB   counterparts, thus  drastically
improving  the position accuracy  to the  sub-arcsec level.   Once the
afterglows   were  identified,   the   full  power   of  optical   and
near-infrared (IR) instrumentation  came  to  play. This  resulted  in  an
overwhelming  diversity of observational  results and  consequently in
the  understanding of  the  properties of  the relativistic  outflows,
their  interaction  with  the  circumsource  medium, as  well  as  the
surrounding  interstellar medium  (ISM)  and the  host galaxies.   The
number  of  well-sampled   optical  and  X-ray  afterglow  lightcurves
increased  rapidly and  revealed a  surprisingly  rich morphology.
Observations  of  re-brightening   and  plateau  phases,  flares,  and
achromatic  or chromatic  breaks  have  become the  norm  and lead  to
adjunctions  to  the  standard   fireball  model,  such  as  multi-jet
components,   extended   central-engine   activity,  and   geometrical
viewing-angle  constraints.  In  addition, a  long-lasting $>100$\,MeV
emission  detected  in  a   handful  of  bursts  by  {\it  Fermi}/LAT
potentially  revealed  an  extension  of the  external  forward  shock
component over eight  decades in photon frequency into the gamma-ray
band \cite{kumar:2010aa}.

Major progress happened in the  field of high-redshift GRBs. While the
highest redshift event at the time  of the $Swift$ launch was GRB 000131
at  $z\sim  4.5$  \cite{Andersen:2000aa},  one year  later  the  $z=6$
barrier was broken with GRB 050904 \cite{Kawai:2006aa}, the QSO record
broken  with GRB  080913 at  $z=6.7$ \cite{Greiner:2008aa},  until GRB
090423  at $z=8.2$  \cite{Tanvir:2009aa,Salvaterra:2009aa} established
the absolute distance record for any  cosmic object for about 1 yr (by
now,     a     galaxy     at     $z=8.56$    has     been     reported
\cite{lehnert10}).  Systematic  near-infrared  observations have  been
recognized as crucial pre-requisite to identify high-z bursts, and two
recently uncovered GRBs with  photometric redshifts in the 9--11 range
promise future progress on the GRB side.

Another field where the rapid GRB coordinate notifications by {\it Swift}
had a particular  impact is that of ``dark  GRBs''.  Originally, those
GRBs with X-ray afterglows  but without optical detection (about 50\%)
were coined  as ``dark GRBs''.  While more refined  classifications of
``dark GRBs''  have been developed,  the possibility of  observing the
afterglow  within  minutes  with  large  ground-based  telescopes  has
increased  the detection  rate  to above  90\%,  thus allowing  secure
statements on the  nature of ``dark GRBs'' to  be made.  Substantially
more  bursts  with $A_{\rm  V}  >0.5$ mag  are  now  revealed than  in
previous samples  \cite{kkz10}, and in many cases  a moderate redshift
(in the  $1<z<4$ range)  enhances the effect  in the observer  frame. The
properties  of this  sample of  early follow-up  demonstrate  that the
darkness can be explained by  a combination of (i) moderate extinction
at  moderate redshift,  and  (ii) an  $\approx$20\%  fraction of  ``dark
GRBs'' at redshift $z>5$ \cite{gkk10}.

The swift dissemination of GRB locations also lead to a significant
increase of high-quality optical afterglow spectroscopy.  Observations
with large ground-based telescopes such as VLT/X-Shooter and
Gemini/GMOS have become standard, and routinely provide redshifts and
detailed views into the structure and chemical composition of the
burst environs.  With the launch of {\it Fermi}, and the exciting new
discoveries with its two instruments LAT and GBM
\cite{Abdo:2009aa,Abdo:2009ab,Abdo:2009ac,Abdo:2010aa,Ackermann:2010aa},
some of the emphasis has moved ``back'' to the prompt emission
characteristics.

\section{Overview of upcoming facilities}

In the next  years before the launch of SVOM  and during its operation
many  new missions  and experiments  will be  brought online  or reach
their full potential.  This list,
presented in Tab.~\ref{tab:facilities}, is
likely not complete, in particular not for  2015 and beyond, since
national programs  for both, ground- as well  as space-based programs,
can  have  a turn-around  time  of  order  5 years.  Thus,  additional
facilities are likely  to emerge, many with strong  synergy with SVOM.
Advancements  are  expected  in   all  areas  of  GRB  research,  from
high-resolution  X-ray spectroscopy,  over  thirty-meter-class optical
and near-infrared  imaging and  sensitive all-sky radio  monitoring to
the   exploration   of  new   cosmic   messengers  (e.g.,   neutrinos,
gravitational waves).

In the following, we describe the expected progress until the SVOM launch
($\S$ 4), during the SVOM mission ($\S$ 5), and end with a few
topics where no progress is obvious, primarily due to missing 
instrumentation ($\S$ 6).

\begin{table}[ht]
\caption{Major ground and space-based facilities expected to be operational
  in 2015 and beyond.}
\renewcommand{\arraystretch}{1}
 \begin{tabular}{llcl}
   \hline
   \noalign{\smallskip}
   Wavelength & Instrument & Start of     &  ~~~~~~Area of impact \\
              &             & Operation    & \\  
   \noalign{\smallskip}
   \hline
   \noalign{\smallskip}
  VHE $\gamma$-rays        & HESS/MAGIC/VERITAS &2003 & prompt emission mechanism; origin of high-energy component\\
             &  HAWC    & 2012 & prompt emission mechanism; origin of high-energy component\\
             &  CTA     & 2016 & prompt emission mechanism; origin of high-energy component\\
   $\gamma$-rays & Fermi    & 2008 & detection and localization\\
             & SVOM     & 2015 & detection and localization; broad-band  afterglow spectroscopy\\
    X-rays   & Swift    & 2004 & detection and localization; broad-band  afterglow spectroscopy\\
             & ASTROSAT  &   2011 & broad-band (UV/optical  to X-ray) afterglow spectroscopy\\
             &   NuSTAR    &   2012& hard-X-ray afterglow spectroscopy \\
             &   eROSITA   &   2013 & detection of orphan afterglows\\
             &   Astro-H   &   2014  & high-resolution X-ray afterglow spectroscopy; chemical composition of environment\\
%    UV       &             &  & \\
     optical & HST         &  1990  & late-time afterglows; host galaxies \\
             & 8-10\,m telescopes& 1999 & all aspects of  afterglows and  host galaxies \\
             &   PanSTARRS &   2009 &  detection of orphan afterglows  \\
             &   Skymapper &   2011  & detection of orphan afterglows   \\
             &   LSST      &   2015 & detection of orphan afterglows \\
             &   GMT/TMT/E-ELT  & 2018 & afterglows of short-hard GRBs; GRB-SNe at $z>0.5$; high-z afterglows and host galaxies\\ 
   (near)-IR   & Herschel & 2009 & host  galaxies \\    
              &   JWST      &   2015 & high-redshift afterglows and host galaxies\\
              &   SASIR     & 2017 & high-redshift afterglows \\
    sub-mm   &   MAMBO    & 2001 & afterglows  and host galaxies\\
             &   SPT      & 2008 & afterglows  and host galaxies \\
             &   SCUBA-2  & 2011 & jet physics; host galaxies\\
             &   ALMA     & 2012 & high-redshift host galaxies\\
    radio    &   LOFAR    & 2010 & detection of orphan and triggered afterglows; jet physics, energetics\\ 
             &   EVLA      &   2010 & afterglow observations; energetics and beaming \\
             & MeerKAT     & 2011  & afterglow observations; energetics and beaming \\
             &   ASKAP       &   2013 & afterglow observations; energetics and beaming\\
    neutrinos  &  ANTARES   & 2008 & explosion physics\\
               & ICECUBE   & 2010 & explosion physics \\
               &  KM3NeT   & 2014 & explosion physics\\
    gravity waves&  Advanced LIGO/VIRGO     & 2015 & detection of gravitational waves from nearby short-hard GRBs \\
      \noalign{\smallskip}
      \hline
\label{tab:facilities}
\end{tabular}
\end{table}

\section{Expected progress until the SVOM launch in 2015}

\subsection{Prompt emission}

The interpretation of the parameters derived from the prompt
emission has long been under debate because of the unknown
correction for the redshift effect. With the advent of the simultaneous
operation of {\it Swift} and {\it Fermi/GBM} there is now a growing
sample of bursts for which both the redshift is measured via the
optical/NIR afterglow, and the prompt emission characteristics
are known beyond the 150 keV upper energy bound of {\it Swift}.
While no results are yet published, the distribution of rest-frame
properties will provide the first solid ground to test the assumptions
made in the modelling over the past four decades.

This sample of common {\it Swift/Fermi} bursts is also expected to 
allow consistency checks on the bulk Lorentz factor $\Gamma$. In the 
gamma-ray regime, both the detection of tens-of-GeV photons as well as
the variability of $>$MeV emission allows to derive lower limits on $\Gamma$.
In the optical, the early, rising part of the afterglow can be
used to infer $\Gamma$. For this method, the redshift to a burst must be known,
and the response after the GRB trigger needs to be fast 
(for $\Gamma\approx1000$, the optical peak is expected only a few sec
after the GRB).

Several bursts have been detected by the {\it Fermi/LAT} 
at energies above 1 GeV, including one short burst. 
Two of the three brightest LAT bursts (090510, 090902B), including the
short burst, showed a clearly separate
second spectral component on top of the usual Band-type simple 
broken power law. 
Whether this component is of leptonic or hadronic origin is widely
debated. One of the most prominent suggestions is a thermal origin in
the photosphere of the fireball \cite{peer2010}.
With the ongoing {\it Fermi} observations it will soon become clear
whether this is a general feature, or occurs only in some bursts.
In either case, it will have a profound impact on our
understanding of how the prompt gamma-ray burst emission
is produced.

% ------------------------------
\begin{figure}[ht]
\includegraphics[width=7.5cm]{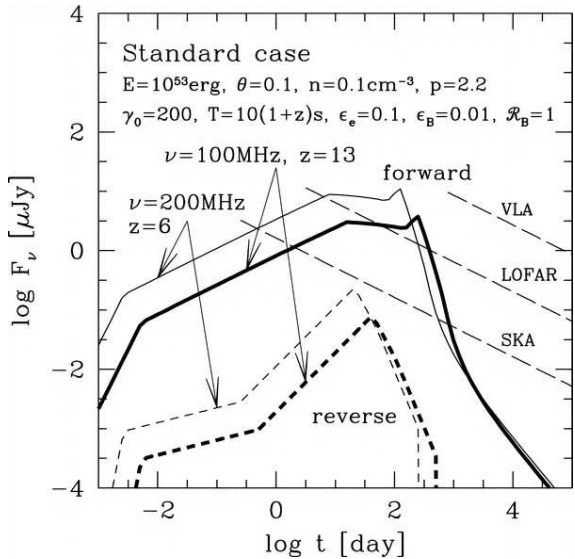}
 \hfill\parbox[t]{6.8cm}{\vspace*{-2.7cm}\caption{GRB afterglow forward
  shock (solid line) fluxes as function of observed time at frequencies
  near the redshifted 21\,cm radiation for two redshifts ($z=6,13$). 
  The 5-$\sigma$ sensitivities for VLA, LOFAR, and SKA are indicated 
  with long-dashed lines (from \cite{Ioka:2005aa}).}
\label{fig:hn_lofar}}
\vspace{0.5cm}
\end{figure}
% ------------------------------

\subsection{GRB jet geometry \& energetics}

The  degree of  collimation  and the  observationally inferred  prompt
emission  and  early   afterglow  energetics  are  tightly  connected.
Collimation  can  be  constrained   in  at  least  two  ways,  through
observations  of achromatic  "jet breaks"  in the  X-ray  and optical
afterglow light curves  and by comparing the number  counts of on-axis
GRBs  and  orphan afterglows.   Only  a  small  fraction of  the  {\it
  Swift}/BAT  detected  bursts has  shown  evidence  for  an X-ray  or
optical jet break \cite{rlb09}.  Here, the general census is that {\it
  Swift}  bursts are  on average  at larger  redshift than  the sample
studied previously.   Thus, their jet breaks occur  typically at later
times in the  observer frame, at flux levels  below the sensitivity of
standard  follow-up campaigns.   Recent  dedicated long-term  afterglow
monitoring with  the more sensitive {\it  Chandra} satellite succeeded
in  recovering jet  break times  for $\approx40$\,\%  of  the observed
events  \cite{bur10}. Further  progress requires  even  more late-time
afterglow observations at optical and X-ray wavelengths.

A direct consequence  of the collimated nature of  GRB outflows is the
prediction of orphan afterglows.   These transients can arise when the
initial  GRB, and its  associated afterglow  light, are  directed away
from  the  observer.   Here,  the  deceleration  of  the  relativistic
outflow, the associated  decrease in special-relativistic beaming, and
the  hydrodynamic  spreading of  the  collimated  jet  combine at  the
jet-break time, $t_{\rm jet}$, to irradiate a fast-increasing fraction
of  the sky \cite{Rhoads:1999aa}.   Observers illuminated  during this
transition  will  detect a  steeply  rising  ($\Delta t\approx  t_{\rm
   jet}/10$)  transient that proceeds  to behave  as a  post-jet break,
optical-X-ray on-axis afterglow.  The substantial uncertainties in the
GRB beaming fraction and  in the redshift and luminosity distributions
make  rate  predictions  for  wide-ange  surveys  difficult.   However,
observations in X-rays with  {\it eROSITA} and optical wavelengths with
e.g., Pan-STARRS, Skymapper, or LSST,  are poised to uncover the first
considerable sample of candidates.

The discovery  of an orphan afterglow  would serve also  as a dramatic
confirmation  of  the  jet  model  for GRBs.   At  radio  frequencies,
afterglows are  bright enough  to be detectable  even years  after the
burst. In particular, the  orphan population detectable with LOFAR and
later SKA will  be dominated by several thousands  of old GRB remnants
(see Fig.~\ref{fig:hn_lofar}).   Such a  large number of  events would
begin to  map out the  beaming distribution and, in  addition, provide
inputs  to physical  models of  relativistic outflows.   The late-time
isotropic radio afterglow emission also  holds the key to deriving the
total burst  energetics. This method has been  successfully applied in
the past, but limited to  the brightest, typically near-by, events due
to  the low  instrumental sensitivities.   The upgrade  of the  VLA to
substantially higher sensitivity (EVLA), and the starting operation of
LOFAR  and ALMA,  the  latter  covering the  peak  of the  synchrotron
spectrum  of   GRB  afterglows,  will  revolutionize   the  field.   A
significant fraction  of, if not all, afterglows  out to high-redshift
will be detected and  calorimetry will provide an unprecedented sample
with well-measured energetics.

\begin{figure}[th]
\includegraphics[width=8.5cm]{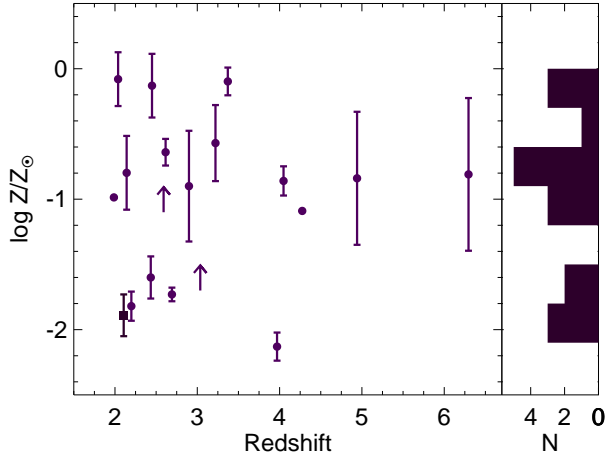}
\hfill\parbox[t]{6.5cm}{\vspace*{-3.5cm}\caption[met]{Left:
metallicities found in GRB DLA systems as function of redshift. 
Filled circles mark metallicities selected from the literature with arrows 
indicating lower limits (adopted from \cite{Savaglio:2009ab}. The filled 
square shows the position of GRB 090926A. The two points without error 
bars are for GRBs 030226 and 050505 for which no uncertainties for 
$N_{\rm H\,\rm{i}}$ have been reported. Right: metallicity histogram 
in bins of $\Delta$log(Z/Z$_{sun}$) = 0.3. 
Lower limits have been excluded here.
From \cite{Rau:2010ab}.
 \label{metal_z}}}
\vspace{0.5cm}
\end{figure}

The  question whether GRBs  can be  used as  standard rulers  has been
discussed since  the discovery of their cosmological  origin. A number
of $\gamma$-ray energy  relationships have been proposed \cite{ama02},
\cite{ggl04}, \cite{ymn04}  and the most promising  indicator has been
identified as an apparently  tight correlation between the peak energy
of   the   GRB   integrated   prompt   emission   spectrum   and   the
collimation-corrected equivalent  energy (so-called Ghirlanda-relation
\cite{ggl04}).  The  currently available  event sample for  which both
quantities are well  constrained is small.  This is  in particular the
result of a poor overlap  between the sample of bursts with accurately
measured prompt-emission spectral parameters (mainly from {\it Fermi})
and the  sample of events with well-localized  afterglows (mainly from
{\it  Swift}) and  jet breaks  and  redshifts. A  slow improvement  is
expected from the  increasing number of bursts detected  by both, {\it
  Swift}  and {\it  Fermi} and  by the  increasing effort  to identify
afterglows  of  bright  {\it  Fermi} GRBs  with  ground-based  optical
wide-field imagers.

\subsection{GRB fireball modelling}

The evolution of the blast wave in the fireball model is governed by
the total energy in the shock, the geometry of the outflow, and the
density structure of the ISM into which it is expanding. The time
dependence of the radiated emission depends on the hydrodynamic evolution
and the distribution of energy between electrons and magnetic field 
\cite{sap99}. 
The  verification  of the  fireball  model  prediction concerning  the
broad-band SED  is severely hampered by the  presently low sensitivity
in the  sub-mm and mm  bands which cover  the peak of  the synchrotron
spectrum.   Indeed, only  a handful  of GRBs  have  measured sub-mm/mm
fluxes  \cite{wig99,gkm08}. Moreover,  the predicted  movement  of the
cooling  break is not  seen in  the majority  of bursts,  despite much
better  coverage   of  the  optical  to  X-ray   regime.   A  dramatic
improvement in the  sensitivity in the sub-mm to  mm range is foreseen
with  the  deployment  of  ALMA  in Chile,  in  particular  after  the
successful commissioning  of the  APEX instrumentation. The  array will
cover the  range from 80 to  720 GHz, with a  predicted sensitivity of
140 $\mu$Jy at 230 GHz and even better at lower frequencies
This should allow the detection and flux monitoring  of a large fraction
of GRB afterglows around the synchrotron peak, and 
in conjunction with the  existing wealth of optical/NIR data, and the
improved sensitivity in radio by EVLA and LOFAR enable
the first systematic observational test of the fireball SED predictions.
Moreover, it will also enable a good characterization of the 
sub-mm properties of many GRB host galaxies. Finally, another 
interesting possibility is that ALMA may be able to measure
the redshift using molecular lines (e.g. CO, [CII] 158 $\mu$m).
[CII] 158 $\mu$m is the main coolant in the Milky Way, is much fainter 
in ULIRGs, and unknown so far in GRB hosts. With the ALMA high-frequency
bands, a redshift range up to 8 will be covered.
%without resorting to optical spectroscopy (not possible anyway at z$>$6). 

\begin{figure}[th]
\centering
\includegraphics[angle=270,width=13.cm]{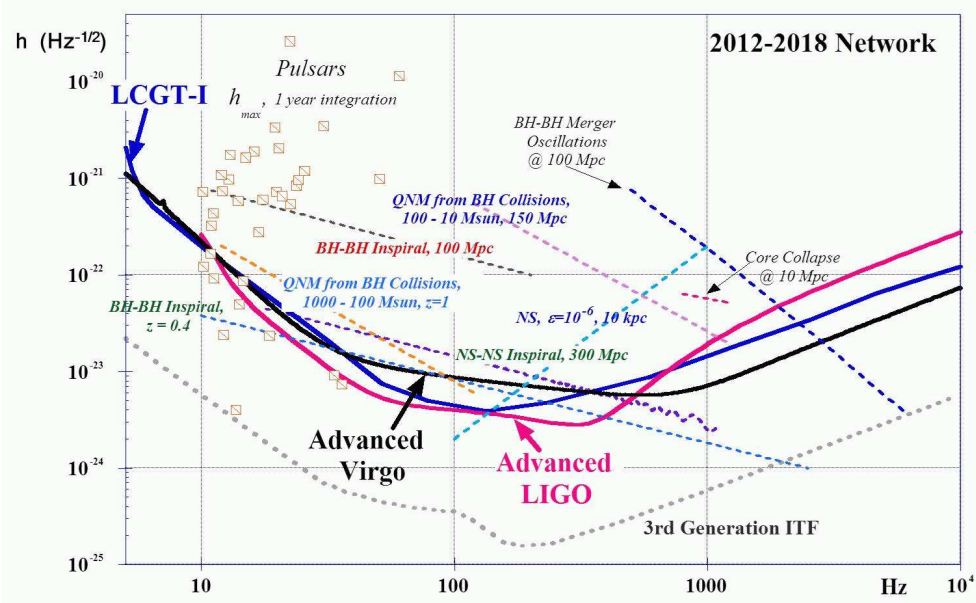}
 \caption[ALIGO]{Sensitivity 
  of ALIGO for compact binary mergers. (From \cite{aligo06})
\label{ALIGO}}
\vspace{0.7cm}
\end{figure}

\subsection{Chemical evolution}

Due  to  their brightness  and  large  distance,  GRB afterglows  are
complementing quasars  as probes of  the chemical
evolution of  the Universe.  Rest-frame  ultra-violet absorption lines
provide important  clues about the  physical state of  the intervening
gas, it's enrichment and dust content, and allow to study directly the
nucleosynthesis processes in massive  star forming regions that host a
GRB.   The  disadvantage of  using  GRB  afterglows  over quasars,
however, is their transient nature.   The fast fading, coupled with an
a priori unconstrained redshift, can easily lead to a limited coverage
of the important spectral  features and diagnostics. In particular the
access  to the  hydrogen  column density  is  often limited due to the 
observational difficulty to measure if for objects at $z<2...2.5$, thus
complicating the abundance estimates.

A  simultaneous   coverage  from   the  atmospheric  cut-off   to  the
near-infrared  has  become  available  with the  installation  of  the
intermediate-resolution ($R=4000-14000$) X-Shooter spectrograph at the
8-m VLT in October 2009.   Its routine use for afterglow spectroscopy
now allows  the detection  of absorption and  emission lines  over the
full  wavelength range accessible  from ground  and promises  to cover
Damped Lyman-$\alpha$ systems  (DLAs) and metallicity diagnostics more
frequently.   This is expected  to lead  to a  steady increase  of the
number  of well-sampled  GRB  sight-lines over  the  next years.   The
existing sample  of $\approx$20 GRB  DLAs (see Fig. \ref{metal_z}) 
should be doubled  until the
launch of  SVOM.  By that time,  in particular the  sample of good-S/N
spectra  of  high-redshift  ($z>4$)  events  will  have  significantly
increased,  and thus enhanced  our understanding  of the  evolution of
metallicity with  redshift.  Previous studies  indicated that contrary
to  quasars  sight-lines,  GRB DLAs  do  not  show  a clear  trend  of
decreasing metallicity            with            redshift
\cite{Savaglio:2009ab,Rau:2010ab}.   However, the  currently available
sample of afterglow DLAs is  dominated by sources in the redshift range
of $2<z<3.5$ with very little coverage beyond that.

The  increasing  number of  good-S/N  spectra  will  also address  the
puzzling  observation that  MgII absorption  systems  intersecting GRB
sight-lines  appear  to  be   stronger  than  their  QSO  counterparts
\cite{Prochter:2006aa,Sudilovsky:2009aa}.  The  answer to whether this
discrepancy lies in  a selection bias or whether  other factors (e.g.,
different  MgII covering  factors,  weak gravitational  lensing of  an
absorber population, or dust  extinction bias) play an important role,
will likely  be concluded from  the significantly increased  sample of
sight-lines.

\section{Expected progress during the SVOM mission (2015 - 2018)}

The  scientific progress  arising directly  from the  SVOM  mission is
described in Chapter~9. Arguably its most important contributions will
be the burst  alerts and the initial localizations  of the optical and
X-ray  afterglows.  At  the point  of  this writing,  it is  uncertain
whether  the  current  burst  flag  ship {\it  Swift}  will  still  be
operational in  2015. Thus,  SVOM may carry  the expectations  for the
whole GRB community for precise locations.  The triggers and 
localizations will be crucial
to maximally exploit the synergy between SVOM and the many experiments
which  will  likely  be   operational  during  its  mission  time,  in
particular  the  new  instrumentation  commissioned  after  2015  (see
Tab. \ref{tab:facilities}).

\begin{figure}[ht]
\includegraphics[width=11.cm]{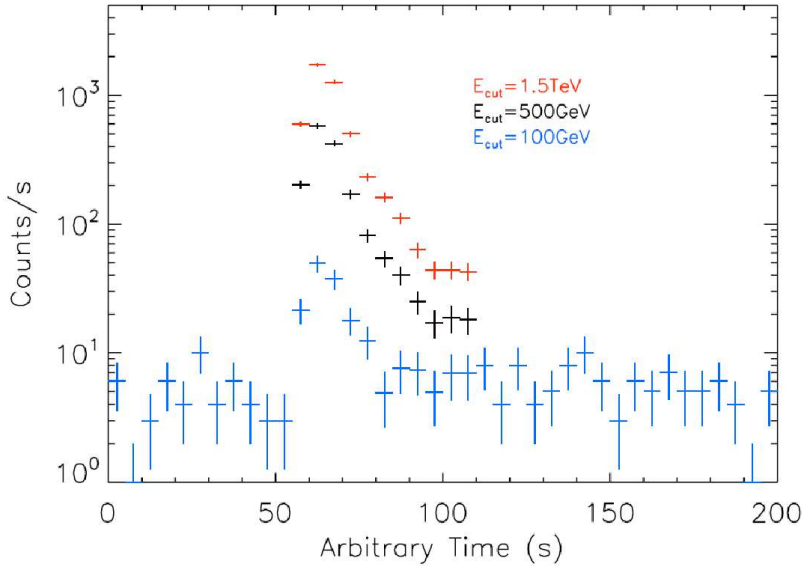}
 \hfill\parbox[t]{4.3cm}{\vspace*{-3.cm}\caption[HAWC]{Simulated 
  HAWC signal for a GRB with a fluence of 
  $1\times10^{-4}$\,ergs cm$^{-2}$, comparable to the 
  10\,keV--10\,GeV {\it Fermi} fluence of GRB~090902B, for 
  three different cut-off energies. (From   http://hawc.umd.edu/science.php)
\label{fig:hawc}}}
\vspace{0.7cm}
\end{figure}

One of  the pressing open  questions in GRB  research is that  for the
energy source of the 'central engine'.  The gamma-rays are produced at
large distances from the  central engine, and carry little information
on the direct energy source.  This is similarly true for the afterglow
photons, though late-time flares have  been used to argue for extended
emission of  the 'central engine'.  At least  for short-duration GRBs,
the detection of  the gravitational wave chirp of  the in-spiral phase
(if the merger  scenario is correct) by an  advanced LIGO+VIRGO system
(Fig.~\ref{ALIGO}) offers  a  unique signature  which will  directly
provide the masses of the two compact objects. Since the gravitational
signal will  also allow to  deduce the luminosity distance,  the prime
energy release can be derived.

For long-duration GRBs,  expectations are a bit more  vague. Events in
the  very nearby  Universe may  produce detectable  gravitational wave
emission from  the associated collapse,  the black hole  formation and
the ring-down phase  \cite{Kobayashi:2003aa}. Similarly, the predicted
neutrino signal  of the  explosion of massive  stars within  the Virgo
cluster  is in  reach of  the  next generation  of neutrino  detectors
(e.g., KM3NeT).  Such bursts are rare,  but if one  could be detected,
the neutrino  signal would  provide a very  good proxy for  the energy
scale  of the  explosion, since  practically all  models  predict that
$\approx$99\% of the energy is carried by neutrinos.
Another aspect of long-duration bursts is the claim, based on a handful
of 'special' bursts, for a population
of nearby, low-luminosity GRBs rather than bursts seen off-axis 
\cite{DaigneMoch07,Foley08}. 
While these bursts are very rare (SVOM might just
detect one or two over its mission lifetime), they are particularly
interesting since the small distance ($z<0.1$) allows detailed studies
of the associated supernova, the host galaxy and the location of the 
GRB therein.

The  {\it Fermi}  satellite  has  produced a  wealth  of new  exciting
results about  the prompt gamma-ray emission (see  Chapter~6). Most of
these,  however, were tightly  linked to  quantities derived  from the
multi-wavelength follow-up, in particular  the redshift. With a likely
overlap  of   both  missions,  SVOM  will  continue   to  provide  precise
localizations for GBM and LAT detected GRBs, thus enabling the crucial
overlap between  prompt and afterglow  measurements.  

\begin{figure}[ht]
\includegraphics[width=7.cm]{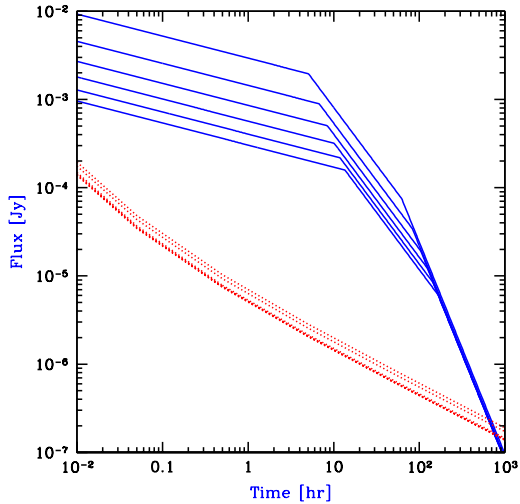}
 \hfill\parbox[t]{7.5cm}{\vspace*{-4.cm}\caption[JWST_GRBspec]{JWST 
  detectability of high-redshift GRB afterglows as a function
of time since the GRB explosion. The GRB
afterglow flux is shown (solid curves) at the redshifted Ly-$\alpha$
wavelength. Also shown (dotted curves) is the detection threshold for
{\it JWST}\, assuming a spectral resolution $R=5000$ with the near
infrared spectrometer, a signal to noise ratio of 5 per spectral
resolution element, and an exposure time equal to $20\%$ of the time
since the GRB explosion. In each set of curves, a sequence of
redshifts is used, $z=5$, 7, 9, 11, 13, and 15, respectively, from top
to bottom. From \cite{Bar04}.
\label{JWST_GRBSpec}}}
\vspace{0.5cm}
\end{figure}

Localizations by  SVOM will also  be important for the  follow-up with
the  next  generation   of  ground-based  very-high  energy  gamma-ray
detectors.     The    High    Altitude    Water    Cherenkov    (HAWC,
\cite{Dingus:2007aa})  experiment  will  surpass  the  sensitivity  of
MILAGRO by more than 15 times between 0.1 and 100\,TeV, while offering
a   similarly   wide  field   of   view   ($2\pi$)   and  duty   cycle
($\approx100$\,\%). These are essential characteristics to measure the
prompt emission of  GRBs (Fig.~\ref{fig:hawc}) and simulations indicate
that HAWC  will be  sensitive to redshifts  of $z\approx1$.   For more
nearby  events, hundreds of  photons above  100\,GeV can  be detected,
allowing HAWC  to directly probe the  bulk Lorentz factor  and size of
the emitting region.  Even higher sensitivity will be reached with the
Cherenkov   Telescope  Array   (CTA,   \cite{Hoffman:2007aa}),  which,
however, will require external event  triggers due to its narrow field
of view.  The current design foresees a high mobility of a part of the
array (180\,deg in  20\,s) allowing CTA to quickly  slew to SVOM burst
locations and to  cover part of the prompt  emission and in particular
the delayed  high-energy component.  GRBs with  measured redshift will
allow to  study the effect of the  optical-IR extragalactic background
light (EBL)  on the shape  of the TeV  spectra of GRBs at  much better
significance. The  EBL directly relates  to the early  star formation,
and  is extremely difficult  to measure  otherwise.  The  present best
constraints  on the  (UV) EBL  are for  GRB 080916C  and  090902B from
Fermi/LAT measurements of GeV emission.

Systematic near-IR follow-up  on bigger ground-based telescopes (e.g.,
SASIR) and from space  (e.g., JWST) promises to significantly increase
the rate of high-redshift identifications. In particular the James Webb
Space Telescope (JWST)  will provide a crucial milestone  in the study
of  the early Universe  through GRBs.   JWST is  an infrared-optimized
6.6\,m space telescope  by NASA with major contributions  from ESA and
CSA.  Its  sensitivity will  allow  detailed  IR  spectroscopy of  GRB
afterglows  even a  week  after the  burst (Fig.  \ref{JWST_GRBSpec}).
This will  be unique in particular  for GRBs at  high redshift ($z>8$)
where ground-based spectroscopy is severely hampered by the sensitivity
of present-day 8--10\,m class telescopes.  It will thus be crucial for
SVOM to  identify high-z GRB  candidates rapidly, in order  to provide
feasible input to  JWST.  With the direct detection  of Pop. III stars
being  impossible with  JWST,  and the  possibility  of detecting  the
supernovae  related  to  the  explosion of  primordial  stars  getting
slimmer, GRBs  presently represent the  best hope of pointing  JWST to
the first star(s).

Another  aspect  where improvements  are  expected  is  the regime  of
high-time resolution optical observations,  both during the prompt GRB
phase as well as in the early afterglow phase.  In the past, high-time
resolution (seconds or faster) optical/NIR follow-up has been obtained
only   occasionally  and  primarily   on  small,   robotic  telescopes
\cite{Stefanescu:2008aa,deCia:2010aa}.   While  variations  on  10-sec
time-scales and  below have  been seen, with  clear deviations  from a
smooth   powerlaw  decay,   attempts  to   improve   the  experimental
capabilities are  very scarce.  High-time  resolution capabilities are
currently  discussed for implementation  in E-ELT  instrumentation and
may have a huge impact in understanding the early afterglow physics.

\section{Areas with no expectation for major progress in the next decade}

We do not anticipate major observational progress on the question of what
role a magnetic field plays in the central engine of GRBs. 
Optical/NIR polarimetry 
facilities exist, but over the last years there has been
little dedication to use them, primarily because the wildly
variable afterglows do not allow clear conclusions to be drawn 
on the possible jet structure and magnetization even if
variable polarisation is detected \cite{gkr03,laz06}.
Similarly,  X- and $\gamma$-ray polarimetry has been proposed, but 
the presently built or proposed small piggyback or balloon instruments do not 
promise a real breakthrough for the next decade,
particularly  time-resolved polarimetry  over the  burst  duration and
comparison  with   the  polarimetric  properties   of  precursors  and
late-time flares.  The mission  GRIPS \cite{gkm10}, proposed to ESA in
the  Cosmic Vision  program, would  be able  to  measure time-resolved
polarimetry  over the  burst duration,  but even  if selected  by ESA,
would not operate before 2020.

Another area of  concern is the simultaneity of  optical/NIR and X-ray
coverage of  the afterglow emission.  During the early  Swift mission,
good optical coverage of  the many exciting X-ray variability patterns
was rare.  With the advent  of dedicated and/or  larger (semi-)robotic
optical telescopes,  the situation has reverted: many  new features in
the optical/NIR  light curves have  no coverage in X-rays.  This could
only  be improved  with an  all-sky  survey mission,  possibly in  the
Lagrange Point 2,  with a large field of view  and sensitivity down to
the sub-mCrab range. Similarly,  more systematic late-time (beyond 3-4
days) X-ray observations would  be useful to investigate the jet-break
issue  as well  as the  nature of  the late-time  optical bumps  - the
corresponding   X-ray   telescopes   are   available,  but   are   not
systematically used so far.

Even  more interesting would  be rapid high-resolution  X-ray spectroscopy  with large
telescopes to  solve the puzzle  concerning the ionized  absorbers and
the WHIM,  or attempt  alternative redshift  measurements. The presently
available {\it Chandra} and  XMM-{\it Newton} telescopes have too slow
slewing times  to allow a  major breakthrough, and this will likely also
be the case for the japanese  Astro-H mission.
 The missions  GRAVITAS and ORIGIN 
(previously EDGE and XENIA)
\cite{xenia}, proposed  to ESA  in the Cosmic  Vision program,  
or DIOS \cite{tub10}, proposed  to the japanese Space Agency, 
are designed to have a  rapid slewing capability, and thus would be 
able to perform these measurements; but even if selected, would
not operate before 2020.

\end{document}